\documentclass[%
superscriptaddress,
preprint,
nofootinbib,
 amsmath,amssymb,
 aps,
]{revtex4-1}
\usepackage{graphicx}% Include figure files
\usepackage{dcolumn}% Align table columns on decimal point
\usepackage{bm}% bold math
\usepackage{color}

\begin{document}

\preprint{APS/123-QED}

\title{Generalized half-center oscillators with short-term synaptic plasticity}% Force line breaks with \\

\author{V. Baruzzi}
\affiliation{Department of Electrical, Electronics and Telecommunication Engineering and Naval Architecture, University of Genoa, 16145 Genoa, Italy}
\author{M. Lodi}
\affiliation{Department of Electrical, Electronics and Telecommunication Engineering and Naval Architecture, University of Genoa, 16145 Genoa, Italy}
\author{M. Storace}
\affiliation{Department of Electrical, Electronics and Telecommunication Engineering and Naval Architecture, University of Genoa, 16145 Genoa, Italy}
\author{A. Shilnikov}
\affiliation{Department of Mathematics and Statistics, \\
 Neuroscience Institute, Georgia State University, Atlanta, GA 30303 USA.}

\date{\today}

\begin{abstract}
How can we develop simple yet realistic models of the small neural circuits known as central pattern generators (CPGs), which contribute to generate complex multi-phase locomotion in living animals? In this paper we introduce a new model (with design criteria) of a generalized half-center oscillator (gHCO), (pools of) neurons reciprocally coupled by fast/slow inhibitory and excitatory synapses, to produce either alternating bursting or synchronous patterns depending on the sensory or other external input. We also show how to calibrate its parameters, based on both physiological and functional criteria and on bifurcation analysis. This model accounts for short-term neuromodulation in a bio-physically plausible way and is a building block to develop more realistic and functionally accurate CPG models. Examples and counterexamples are used to point out the generality and effectiveness of our design approach.      
\end{abstract}
\maketitle

\section{introduction}
Central pattern generators (CPGs) are small neural circuits that can autonomously (i.e., in the absence of sensory feedback or higher motor planning centers inputs) produce various rhythmic patterns of neural activity  \cite{harris2017neural}. They bear a fundamental function in both invertebrate and vertebrate animals as they determine multi-phase locomotion -- the innate motor behavior that requires sequential activation of body muscles in a coordinated way \cite{kiehn2016locomotion}.  Various approaches to the modeling of CPGs and CPG-inspired control systems have been explored in the last decades \cite{buono2001models,pinto2006central,ijspeert2008central,yu2013survey,danner2017computational}. Recently, new methods have been proposed to reduce large models of detailed neural networks to smaller CPG circuits, trading off biological plausibility and complexity of the model \cite{buono2001models,pinto2006central,molkov2015mechanisms,lodi2017design,ausborn2018state}.\\
\noindent 
Although CPGs function autonomously, their activity is modulated through the influence of hierarchically higher areas, which can, for example, prompt transitions between gaits \cite{grillner2006biological,takakusaki2013neurophysiology,caggiano2018midbrain}.  A single gait in a typical CPG model is obtained by fixing the connectivity. By contrast, to generate multiple gaits the CPG connections between constituent neurons are typically changed acting on the synaptic weights to model the control action of the brainstem \cite{danner2017computational,molkov2015mechanisms,lodi2017design,lodi2019design}. The modulation from higher areas that controls the synchronization between the CPG neurons, and thus triggers gait switches, is conveniently integrated in CPG models to directly affect the synaptic conductance strengths. However, in real CPGs changes in conductance values are the result of long-term synaptic plasticity, and therefore it is hardly a cause for quick gait switches, which can instead be accounted for more realistically by short-term neuromodulation.
Indeed, most natural CPGs exhibit patterns of functional connectivity between neurons or synchronized clusters of neurons that can undergo spontaneous fluctuations and be highly responsive to perturbations, e.g., induced by sensory input or cognitive tasks, on a timescale of milliseconds or hundreds of milliseconds, respectively, thus ensuring robustness and stability. This short-term neuromodulation lacks in most CPG models.\\
\noindent 
One of the pivotal building blocks of many CPGs is a half-center oscillator (HCO). The HCO-concept is widely used to model two synchronous pools of neurons reciprocally inhibiting each other to produce stable rhythmic alternation in animal locomotion \cite{brown1914nature,calabrese1995half}. This basic structure has been largely studied from both biological and nonlinear dynamics standpoints. For example, in \cite{bem2004short} transitions between stable synchronous states in the HCO occur through direct manipulations with synaptic weights, whereas in \cite{doloc2011database} a large database of HCO models is swept using a brute-force approach, without a focus on gait transitions. While the importance of an interplay between inhibitory and excitatory coupling has already been outlined \cite{bem2004short}, the thorough understanding of its functional role for determining multiple states or patterns in such neural networks and how transitions between them may stably occur remains yet insufficient.
Moreover, there is the growing evidence that (i) post-synaptic potential (PSP) summation increasing with the spike frequency in the pre-synaptic cell is a crucial factor for stable functioning of some CPGs \cite{dale1985dual,pinco1994synaptic,Sakurai2017,Sakurai6460}, while other experiments indicate that (ii) the activity of some synapses is barely affected by the spike frequency \cite{danner2017computational}. \\
\noindent 
In this paper, we propose a generalized half-center oscillator (gHCO) composed of two neurons or of two neural pools that are coupled reciprocally by excitatory synapses, in addition to the standard HCO's reciprocally inhibitory synapses. We show that this circuitry warrants a more biologically plausible mechanism of short-term plasticity to implicitly control the phase-lag between the gHCO cells by varying their spike frequency through sensory drive or external currents, rather then directly manipulating the synaptic conductance strengths. Moreover, we show how to calibrate the gHCO parameters in order to obtain the desired behaviors, also carrying out a numerical bifurcation analysis.

\section{The gHCO and its design constraints} 
\begin{figure}[!b]
	\includegraphics[width=0.5\textwidth]{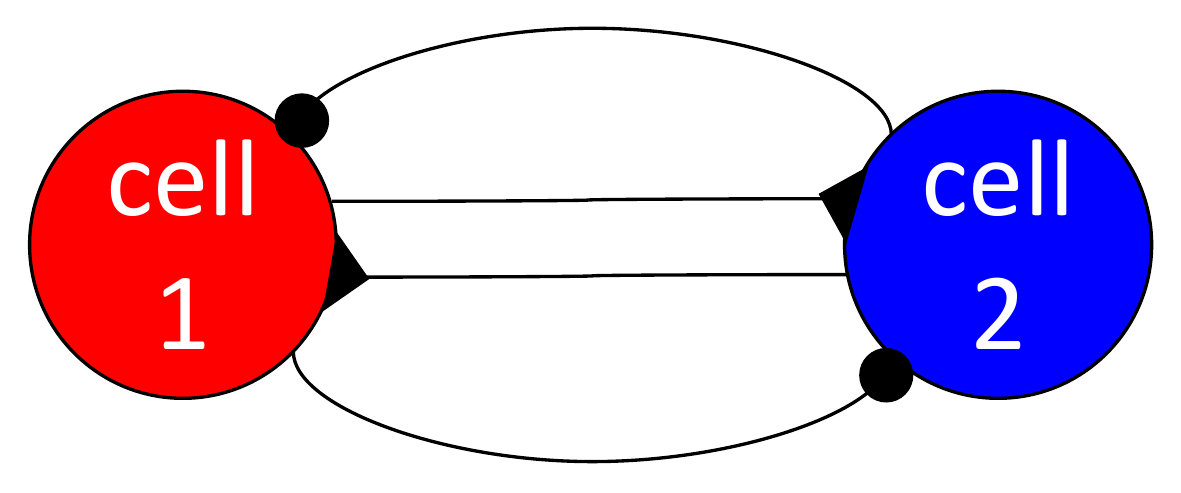}
	\caption{\label{fig:HCO}(color online). gHCO neural circuit with inhibitory (denoted with $\blacktriangleright$ $\blacktriangleleft$ ) and excitatory ($\bullet$) synapses reciprocally coupling two oscillatory cells.}
\end{figure}

The proposed generalized half-center oscillator is shown in Fig.~\ref{fig:HCO}. It is made of two neurons or two neural pools, coupled by both excitatory (marked by a black circle) and inhibitory (marked by a black triangle) synapses.

\begin{figure}[!b]
	\includegraphics[width=1\textwidth]{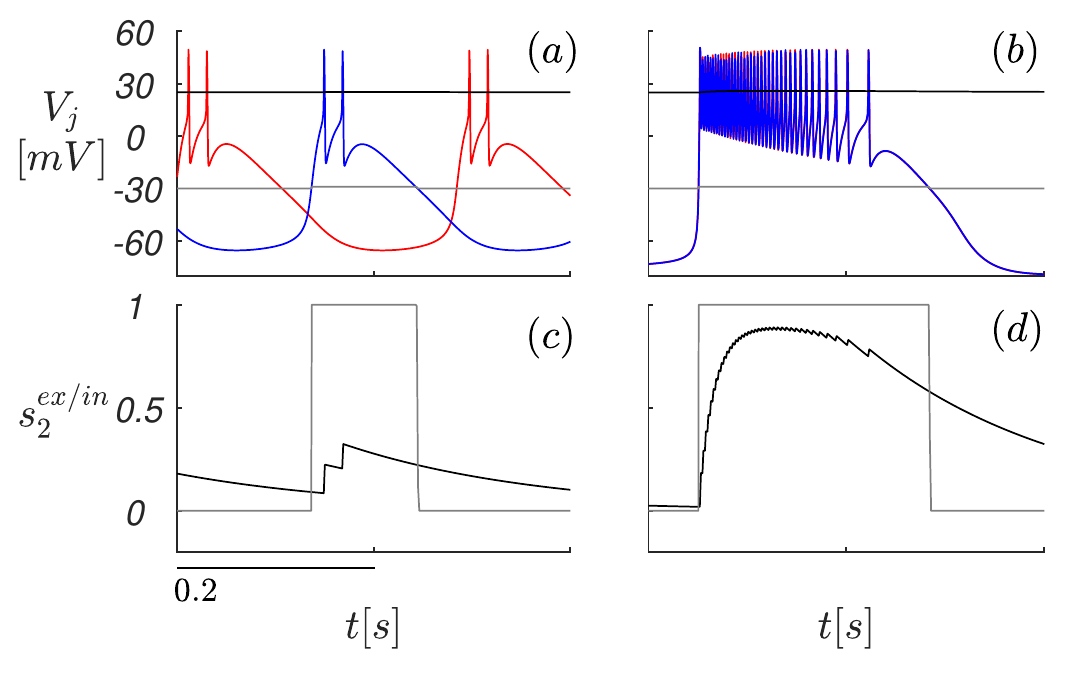}
	\caption{\label{fig:exc_inh_act} (color online) Asymptotic anti-phase (a) and synchronous (b) bursting voltage traces $V_1$ (red) and $V_2$ (blue) at $I_{c}=-0.43$ and 0.13, resp., in gHCO~(1)-(3), superimposed with excitatory/inhibitory thresholds $\theta^{}$ (horizontal lines) at 10 and -30~mV. (c,\,d) Synapse dynamics: fast modulatory $s_{2}^{in}(t)$ (gray) vs. slowly summating/decaying $s_{2}^{ex}(t)$ (black). See the Appendix for parameters.}
\end{figure}

There are a few simple constraints that neurons and synapses must meet for the circuit to generate stably the desired rhythmic outcomes: (a) both  neurons are endogenous bursters with (b) the spiking voltage range above the hyperpolarized voltage (i.e., they do not \textit{undershoot} \cite{izhikevich2000neural}) within each burst, while (c) the mean spike frequency can be controlled. The gHCO bursters are coupled by (d) slow synapses with PSP summation whose strength increases with the growing spike frequency in presynaptic cells, as well as by (e) fast synapses without PSP summation.\\ 
\noindent 
In what follows, both gHCO cells are represented by the Hodgkin-Huxley (HH) type model of the thalamic reticular neuron \cite{destexhe1994model,nagornov2016mixed} (see Appendix). This slow-fast model with seven state variables can exhibit endogenous bursting activity of alternating trains of fast action potentials with long quiescent intervals, as depicted in Fig. \ref{fig:exc_inh_act}. The dynamics of the membrane potential $V_j$ and of the voltage-dependent state variables (the vector $\mathbf{y}_j$) are governed by a generic set of HH-like equations 
\begin{equation}
    \label{eq:reticular}
    \frac{d}{dt}\begin{bmatrix}
           V_j \\
           \mathbf{y}_j \\
         \end{bmatrix} = \begin{bmatrix}
           - \sum_k I_{k} + I_j^{syn}\\
           f(V_j,\mathbf{y}_j) \\ 
         \end{bmatrix}, \quad \mbox{where}~~ j = 1,2. 
\end{equation}
% $f_1: \mathbb{R}^n \rightarrow \mathbb{R}$ and $f_2: \mathbb{R}^n \rightarrow \mathbb{R}^{n-1}$.
where $f(V_j,\mathbf{y}_j)$ is a vector function describing $\mathbf{y}_j$-dynamics; in particular, each $f$ component for the HH gating variables is a logistic function.
In addition to intracellular currents, $\sum_k I_{k}$ includes a further external contribution, namely a control current $I_{c}$ acting essentially on the spike frequency within bursts.  For the given model, bursting activity occurs when $I_{c} \in  [-0.43, \, 0.13]$ $\left [\frac{\mu A}{cm^2} \right]$, with the mean spike frequency decreasing from 15.36 to 4.13 ms. 
The term $I_j^{syn}$ is the incoming mixed, excitatory/inhibitory synaptic current originating from the $i$-th cell onto the $j$-th, post-synaptic cell: 
\begin{equation}
    \label{eq:Isyn}
    I_{j}^{syn} = g^{ex}(E^{ex}-V_j)s^{ex}_{i}+g^{in}(E^{in}-V_j)s^{in}_{i},
\end{equation}
where $E^{ex/in}$ are the reversal potentials for excitatory/inhibitory synapses and $ 0 \le s_i^{ex/in} \le 1$ is the activation or neurotransmitter release rate of the synapse, excitatory ($V_j < E^{ex}$) or inhibitory ($V_j > E^{in}$). For the slow synapses with PSP summation we employ a first-order dynamic synapse \cite{wang1999fast,buonomano2000decoding,jalil2012spikes}.
% a newly developed model \cite{scully2019private} of an {\em inflection} $\alpha$ synapse \cite{destexhe1994synthesis,jalil2012spikes}.
The dynamic evolution of its activation rate is governed by the following equation
\begin{equation}
    \label{eq:alpha}
    \frac{ds_{i}}{dt} = \alpha\,(1-s_{i})f_{\infty} (V_i)-\beta s_{i},\quad 
    f_{\infty}=\frac{1}{1+e^{-\nu(V_i-\theta)}},
\end{equation}
where $\theta$ is the synaptic threshold, whereas $\alpha$ and $\beta$ are coefficients weighting the raise and decay terms, respectively. 
% The key features of this synapse is that its mean activation rate $\langle s_i \rangle $ remains small for low spike frequency values while its dynamic maximization and the further saturation at greater frequencies has a characteristic inflection point of a low-band filter \cite{Lindner2076,Reinartz}, for proper choices of $\alpha$ and $\beta$ matching the temporal properties of the given neuron. 
To model the static synapses without PSP summation we employ the fast threshold modulation paradigm \cite{somers1993rapid} using the sigmoidal function: $ 0 \le s_i=f_{\infty}(V_i) \le 1$, with $\theta$ being below the spike-level. \\
\noindent 
To illustrate the contrasting properties of these synapse models, we refer to Fig.~2, showing the bursting voltage traces $V_1$ (red) and $V_2$ (blue) and the synaptic activation dynamics, fast $s^{in}_{2}(t)$ (gray) and slow $s^{ex}_{2}(t)$ (black) at the edge of the $I_{c}$ bursting interval. Observe that the neurotransmitter release rate $s^{in}_{2}(t)$ of the fast FTM synapse (1) is maximized as soon as the voltage $V_2(t)$ in the pre-synaptic cell overcomes the synaptic threshold $\theta^{in}$ (indicated by the grey lines in panels a,b), (2) remains constant regardless of the spike frequency, and (3) vanishes with the burst termination. In contrast, the low spike frequency (panels a,c) barely activates the slow synapse (see $s^{ex}_{2}(t)$) that at high spike frequency (panels b,d) exhibits the profound PSP build up; the ascending rate is ruled by $\alpha>0$, and the exponential decay due to $\beta>0$ starts after the voltage lowers below $\theta$.  

\section{Parameter calibration}
The neuron and synapse models~(1-3) are calibrated to physiologically plausible values to meet the above requirements (a)-(e) and to ensure a smooth  and reversible  transition from anti-phase to in-phase bursting occurring in the gHCO as the spike frequency changes due to $I_{c}$-variations. Just to clarify things, let us consider the dynamics of the gHCO with fast FTM inhibitory and slow excitatory synapses. Moreover, the corresponding synaptic thresholds are set at $\theta^{in}=-30$ and $\theta^{ex}=25$mV, respectively. As such, the inhibitory synapses without PSP summation (de)-activate quickly and their strength remain constant during each burst regardless of the spike frequency. In contrast, the slow excitatory synapses exhibit PSP summation that becomes stronger with an increase of the spike frequency.

\noindent 
Figure 2 shows that at the low end $I_{c}=-0.43$ of the bursting region, near the transition to the hyperpolarized quiescence, the gHCO neurons oscillate in anti-phase with the smallest number of spikes per burst and lowest spike frequency (panels (a, c)), whereas on the opposite side at $I_{c}=0.13$ the neurons burst in phase with a larger number of spikes per burst and with much higher spike frequency (panels (b, d)). Changing the value of $I_{c}$ changes the strength of the excitatory synapses, and hence the proportion between inhibition and excitation that repel the gHCO neurons or attract them to each other, respectively.  The {\em  
phase-lag} $\Delta$ (defined on mod~1) between burst initiations in the neurons \cite{jalil2013toward,wojcik2014key,zhao2015experimental} allows quantifying the phase-locked states produced by the gHCO. In case of the synchronous or in-phase bursters, $\Delta = 0$ (or $\Delta = 1$). When they burst in alternation, with $\Delta = 0.5$, we say that they are in anti-phase. The intermediate values of $\Delta$ correspond to ``winner-less'' patterns transitional between the in- and anti-phase states generated by the gHCO. 
\begin{figure}[!b]
\includegraphics[width=1\textwidth]{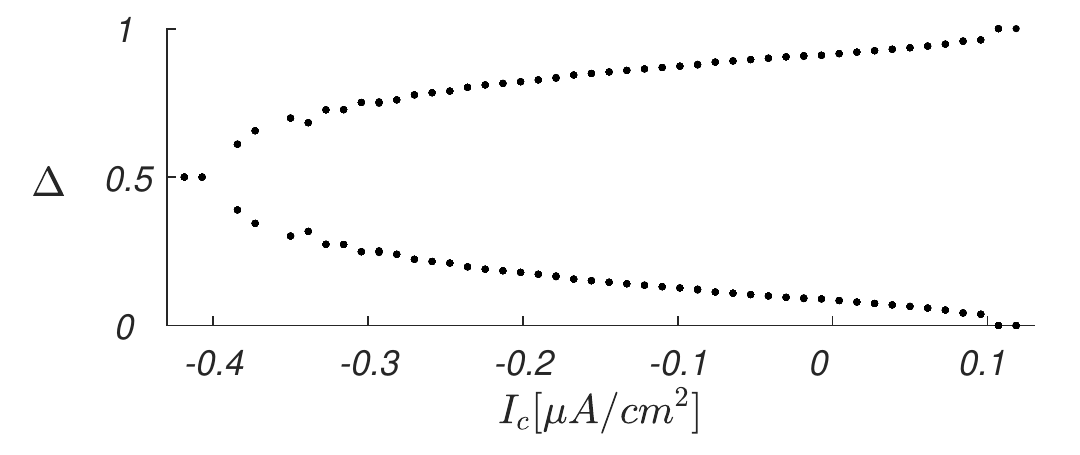}
\caption{\label{fig:bif_diag_s-alpha} Bifurcation diagram showing how the phase-lag $\Delta$ between the gHCO neurons is affected by the current $I_{c}$; here, 30 initial $\Delta$-values were sampled evenly between 0.05 and 0.95 for each of the 50 $I_{c}$-values. Parameters listed in the Appendix.}
\end{figure}

The bifurcation analysis of the system~(1--3) was carried out using the computational toolbox CEPAGE \cite{lodi2017cepage}. Since we want the gHCO to transition from anti-phase regime to in-phase regime varying $I_{c}$, we need the proportion between inhibition and excitation to be significantly different for the two values of $I_{c}$ at the edges of its range. To this end, we seek maximum difference in the mean values of $s_{i}^{ex}$ (over one period) at the two extreme values of $I_{c}$, i.e., -0.43 (anti-phase pattern) and 0.13 (in-phase bursting).
% and we set the values of $\alpha$, $\beta$ and $\theta^{ex}$ accordingly (see Supplemental Material).
We set the numerical values of $\theta^{ex}$, $\alpha$ and $\beta$ according to this principle, running a set of simulations over a grid of parameter values: $\theta^{ex}=\{10, 25\}$, 10 evenly spaced values of $\alpha \in [0.05, 1]$  and 10 evenly spaced values of $\beta \in [0.005, 0.1]$. The considered values of $\theta^{ex}$ indicate voltage levels representative of two different conditions: at $\theta^{ex}=10$ each spike appears broader, i.e. $V_j$ stays above $\theta^{ex}$ for a longer time window; at $\theta^{ex}=25$ each spike appears narrower, i.e. $V_j$ stays above $\theta^{ex}$ for a shorter time period. We choose the parameter setting that provides maximum difference in the mean values of $s_{i}^{ex}$ for the two extreme values of $I_{c}$ (see Appendix B).
The synaptic conductances $g^{in/ex}$ are set to obtain anti-phase synchronization for low spike frequency, condition in which the mean value of $s_{i}^{ex}$ is minimum, and in-phase synchronization for high spike frequency, condition in which the mean value of $s_{i}^{ex}$ is maximum.

The results are summarized in Fig. \ref{fig:bif_diag_s-alpha}, and reveal the dependence of the phase-lag $\Delta$ on the $I_c$-current, and hence explicitly on the spike frequency within bursts. As expected, at low $I_c$-values between $-0.43$ and $-0.40$, the fast reciprocal inhibition within the gHCO dominates and makes its neurons burst in alternation with $\Delta = 0.5$. As the $I_c$-current is increased, the spike frequency raises, which in turn makes the slow excitatory synapses sum up faster and stronger on average. With larger $I_c$ values, the reciprocal excitation gradually prevails over the reciprocal inhibition, which gives rise to the smooth onset of the stable in-phase bursting in the gHCO. This is revealed in the bifurcation diagram with a characteristic pitchfork shape of the dependence of the phase-lag $\Delta$ on the $I_c$-current. 
We note also that this diagram has been obtained by making a multi-shooting for each parameter value. This is a direct indication that there is no hysteresis and therefore the absence of multi-stability or the coexistence of anti- and in-phase bursting for same parameter values, and that the transition between activity rhythms is continuous and reversible.
% We also note that both the increase and decrease of the $I_c$-current result in almost the same diagram which a small overlap in the given scale. Such an overlap is indicative of a hysteresis due to bistability or the coexistence of anti- and in-phase bursting for same parameter values. Other than that, the transition between activity rhythms is nearly continuous and reversible.
We would like to re-emphasize that the maximal synaptic conductances $g^{in/ex}$ in Eq.~(2) once set are not changed, and the transition is solely determined by the gradual increase/decrease of the mean $s_{i}^{ex}$-value caused by the spike frequency variations in the gHCO neurons. 

\begin{figure}[!b]
	\includegraphics[width=1\textwidth]{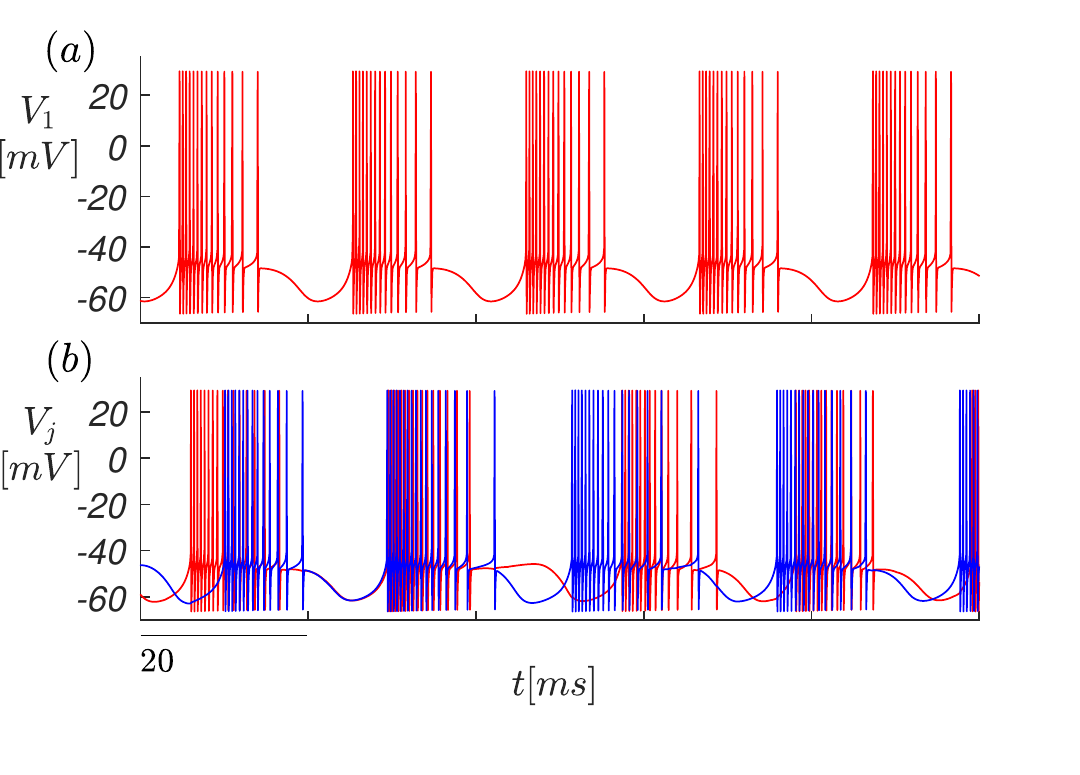}
	\caption{\label{fig:condition_B} (color online). (a) Asymptotic bursting voltage trace with undershoot produced by the Plant neuron model \cite{plant1981bifurcation,alaccam2015making}. (b) Voltage traces produced by the gHCO with two coupled Plant neurons. See the Appendix for parameters.}
\end{figure}

\section{Counterexamples}
\begin{figure}[!t]
	\includegraphics[width=1\textwidth]{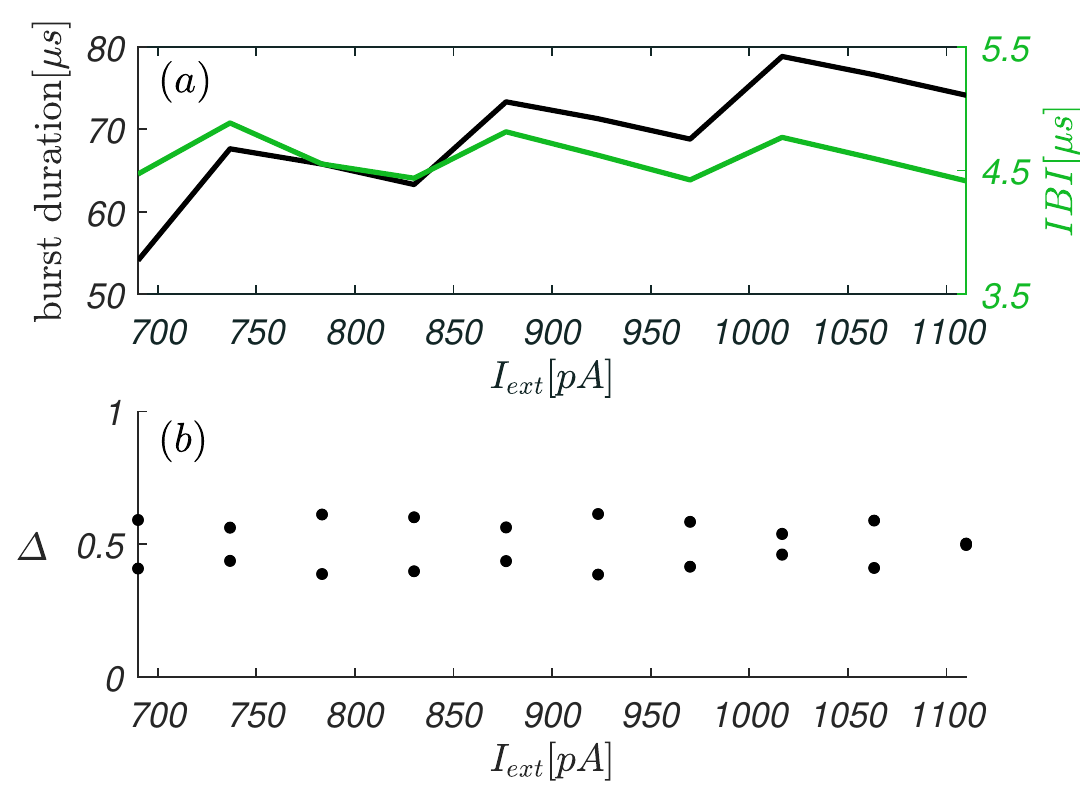}
	\caption{\label{fig:condition_C} (color online). (a) Mean values (over 5 s) of the IBI (green line) and the burst duration (black line)  plotted against \(I_{ext}\) for the exponential IF-model \cite{brette2005adaptive}. Corresponding bifurcation diagram for the phase lag $\Delta$ between the cells in the gHCO, in which each cell is an exponential IF-model (b).}
\end{figure}

The proposed gHCO concept can fall apart whenever one or more of the conditions on the neuron and synapse models are not fulfilled. If the bursting condition (a) is broken, the approach is no longer applicable. Two neurons, spiking in isolation, can burst in alternation due to reciprocal inhibition, but not through reciprocal excitation, which makes both even more synchronously depolarized with a higher frequency. 
If the neurons undershoot (condition (b)), which is typical for elliptical bursters \cite{alaccam2015making} (see Fig. \ref{fig:condition_B}(a)), the choice of the inhibitory threshold $\theta^{in}$ to warrant evenly constant activation $s_{i}^{in}$ requires additional considerations. Indeed, this choice can result in less robust dynamics of the gHCO, due to inhibition-excitation competition (see Fig.~\ref{fig:condition_B}(b)).
Condition (c), outlining the importance of being able to control spike-frequency and not only burst duration of the pre-synaptic cell,
% outlining that $s_{i}^{in}$-build up is more spike-frequency dependent then is affected by the burst duration
is quite crucial for stable gHCO functions. To point out its significance, we employ the exponential integrate-and-fire (eIF) neuron model \cite{brette2005adaptive}, where an external current $I_{ext}$ primarily controls the burst duration with insignificant spike-frequency variations, as shown in Fig. \ref{fig:condition_C}(a). In this scenario, the activation of both inhibitory and excitatory synapses is mainly determined by the burst duration in the eIF-neurons, and thus $I_{ext}$-variations can only cause proportional changes in the average excitatory  $s_{i}^{ex}$- and inhibitory $s_{i}^{in}$-values. As a result, neither inhibition nor excitation can solely dominate and produce the expected solo stable anti-phase or in-phase bursting patterns within the given $I_{ext}$-range, as shown in Fig.\ref{fig:condition_C}(b). Conversely, changing the parameter $g_e$ of the eIF neuron model significantly modifies the spike frequency, and the corresponding bifurcation diagram has the characteristic pitchfork shape, as expected. However, the parameter $g_e$ is a conductance, and thus is not a realistic control parameter, according to our guidelines.
\begin{figure}[!t]
	\includegraphics[width=1\textwidth]{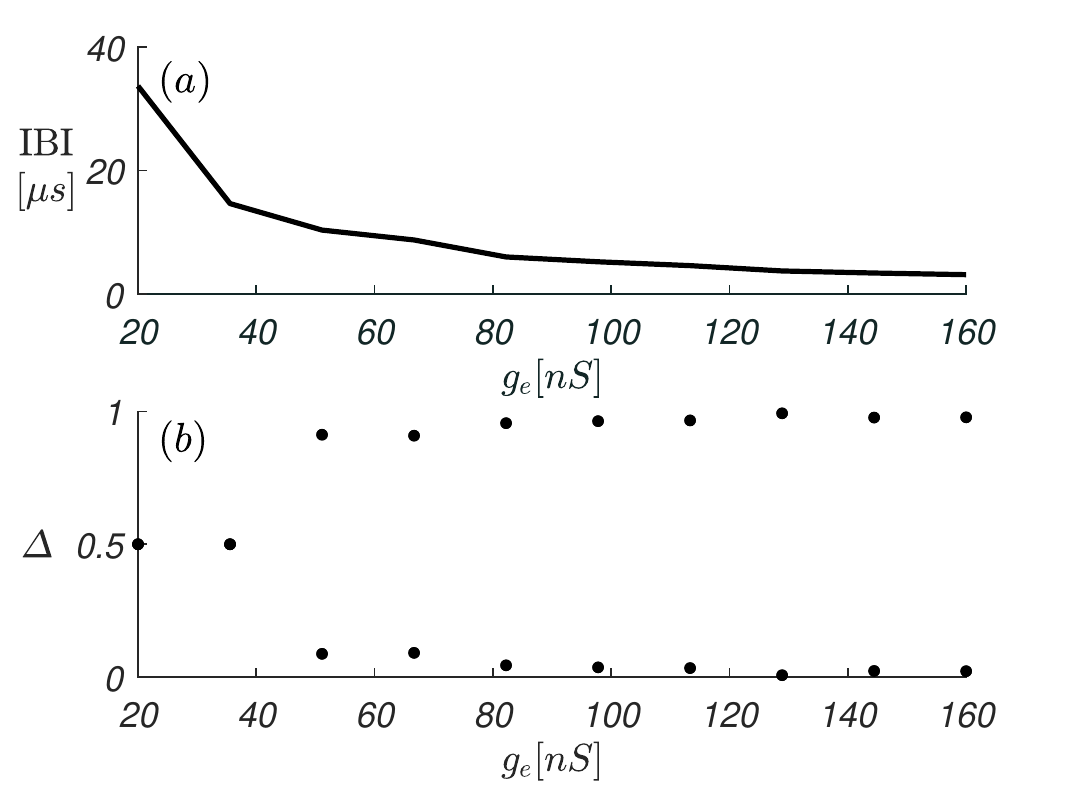}
	\caption{\label{fig:support} (a) Mean values (over 5 s) of the IBI plotted against \(g_e\) for the exponential IF-model \cite{brette2005adaptive}. Corresponding bifurcation diagram for the phase lag $\Delta$ between the cells in the gHCO, in which each cell is an exponential IF-model (b).}
\end{figure}
\noindent 
Condition (d) follows (c), as the synaptic threshold $\theta$, for the slow synapses, has to be within the spike voltage range of the pre-synaptic neuron and the dynamics is to be slow enough to allow $s_i(t)$ to grow and the synapse to exhibit PSP summation.
% Moreover, if the dynamics is too fast, the activation drops to zero between two consecutive spikes of the burst. This does not allow the activation to increase during the burst of the pre-synaptic neuron even when the IBI is low.
Condition (e) guarantees that the activation of the fast synapse does not exhibit PSP summation and hence does not change due to spike frequency variations in the pre-synaptic neuron. 

\begin{figure}[!b]
\includegraphics[width=1\textwidth]{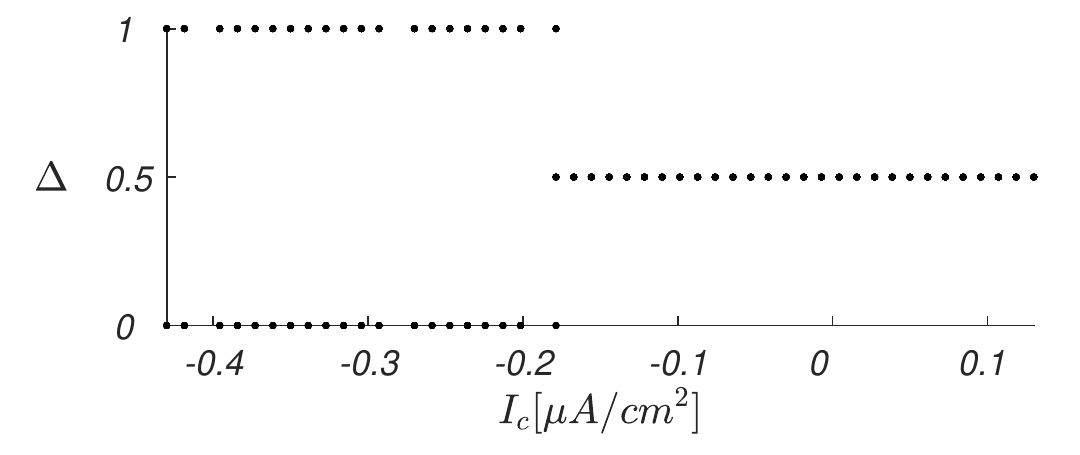}
\caption{\label{fig:bifurcation_swap} Bifurcation diagram showing the flat-even phase-lags, $\Delta=\{0,\,1\}$ (in-phase) and $\Delta=0.5$ (anti-phase), between the bursters plotted against the current $I_{c}$ for the gHCO with slow inhibitory and fast excitatory synapses; here, 30 initial $\Delta$-values were sampled evenly between 0.05 and 0.95 for every $I_{c}$ value out of 50. Parameters listed in the Appendix.}
\end{figure}

\begin{figure}[!b]
\includegraphics[width=1\textwidth]{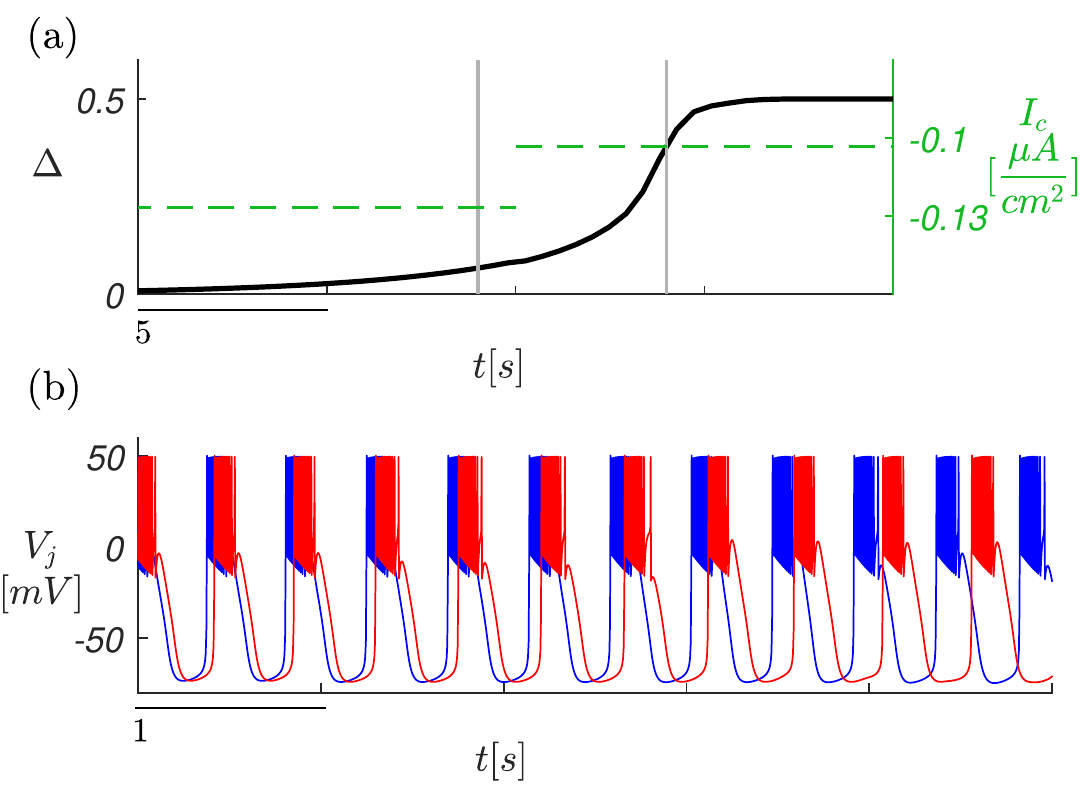}
\caption{\label{fig:smooth} (color online) (a) Time evolution of the phase lag $\Delta$ between the gHCO cells (black line) in response to step-wise changes of $I_{c}$ (green dashed lines); $I_{c}$ increased over 25 steps from $-0.43$ to $0.13$, only the time window in which $\Delta$ transition occurs is shown.
(b) Voltage traces progressing from in-phase to anti-phase bursting within the time window bounded by the grey vertical lines in (a).}
\end{figure}

\section{Towards a locomotion CPG}
As the gHCO often happens to be a CPG building block, we discuss some solutions ensuring that both the phase lags and the burst frequency are consistent for the modeled gaits. 
For instance, in left-right alternation of the mouse locomotion, a phase lag $\Delta = 0.5$ occurs at low burst frequencies (walk and trot gaits), whereas a phase lag $\Delta$ close to 0 (or to 1, equivalently) occurs at high burst frequencies (gallop and bound gaits) \cite{bellardita2015phenotypic,lemieux2016speed,lodi2019design}.
Recall that the thalamic reticular neuron model in isolation exhibits high frequency bursting at small $I_{c}$-values and slow bursting at greater $I_{c}$-values. Therefore, for the gHCO built with such models to produce in-phase/anti-phase synchronization at high/low burst frequencies for the desired gaits, the time-scale of the synapses in its circuitry should be swapped: slow inhibitory synapses with PSP summation and FTM-fast excitatory ones without PSP summation, see the Appendix for details. Moreover, we use a modified version of the first-order synapse to model slow inhibitory synapses.
The dynamics of its activation is governed by the following equation
\begin{equation}
    \label{eq:alpha2}
    \frac{ds_{i}}{dt} = \alpha\,s_{i}\,(1-s_{i})f_{\infty} (V_i)-\beta s_{i}
\end{equation}
where the new multiplicative term delays and hence slows down the synaptic activation for low spike-frequency in the pre-synaptic neuron; the synapse remains inactive near $I_{c}=-0.43$. 
The synapse given by Eq. 4 maintains a greater contrast in the mean $s_i$-values corresponding to the low and high ends of the bursting $I_{c}$-range for the given neuron model.
The results are summarized in Fig. \ref{fig:bifurcation_swap}, representing the bifurcation diagram for this gHCO. It demonstrates that the gHCO bursters oscillate robustly  in-phase ($\Delta=\{0,\, 1\}$) for negative $I_c$-values and rapidly transition to the stable anti-phase ($\Delta=0.5$), phase-locked state  as the drive is increased above -0.2. Despite the  abrupt jump in the bifurcation diagram, the time evolution between in-phase and anti-phase bursting occurs smoothly (see Fig.~\ref{fig:smooth}) as the control current $I_c$ is step-wise increased from -0.43 to 0.13.

\noindent 
\section{Concluding remarks}
We developed a generalized HCO-model with a short-term plasticity mechanism, which accounts for short timescale gait transitions induced by sensory input or cognitive tasks. The proposed concept is based on simple constraints  (i)  subjecting models for cells and synapses and (ii) optimizing the trading-off between physiological plausibility and model functionality. The generality of our approach suggests that it will be applicable for other biologically plausible and phenomenological models of endogenous (square-wave) bursters, and for other dynamic synapse models.  
%   \ms{We could point out that there are other models of neurons and synapses that work well, showing some results in the supplementary material. We could use the IF neuron model (acting on $g_{exp}$) with the dynamic alpha synapse and/or the reticular neuron with the dynamic alpha synapse.}

%Breaking the necessary conditions leads to unsuitable gHCOs, as shown through the proposed counterexamples.
\vspace{0.4cm}
\begin{acknowledgments}
We would like to acknowledge J.~Scully's contribution to the concept and development of the synapse model Eq. (4). A.S.'s research was partially funded by the NSF grant IOS-1455527. M.S. and A.S. conceptualized the work; V.B. and M.L. conducted the experiments.
\end{acknowledgments}

\newpage

\appendix
\section{Neuron Models}
\subsection{Thalamic Reticular Neuron Model}
The thalamic reticular neuron model \cite{destexhe1994model,nagornov2016mixed} is defined by the following state equations:
% \begin{equation}
% \begin{array}{l}
% \dfrac{dV}{dt} = \frac{-I_T-I_L-I_{Na}-I_{K}-I_c+I^{syn}}{C}, \\\vspace{5pt}
% I_T = g_{Ca}m_T^2h_T(V-E_{Ca}), \quad I_L = g_L(V-E_L), \\
% I_{Na} = g_{Na}m^3h(V-E_{Na}), \quad I_K = g_kn^4(V-E_k), \\

% \dfrac{dh}{dt} = 0.128e^{\frac{17-V}{18}}(1-h)-\frac{4}{e^{-0.2(V-40)}+1}h, \\\vspace{5pt}

% \dfrac{dm}{dt} = \dfrac{0.32(13-V)}{e^{0.25(13-V)}-1}(1-m)-\frac{0.28(V-40)}{e^{0.2(V-40)}-1}m, \\\vspace{5pt}

% \dfrac{dn}{dt} = \dfrac{0.032(15-V)}{e^{0.2(15-V)}-1}(1-n)-0.5e^{\frac{10-V}{40}}n, \\\vspace{5pt}

% \dfrac{dm_T}{dt} = \frac{m_T^\infty-m_{T}}{\tau_{mT}}, \quad m_{T}^\infty = \frac{1}{1+e^{-\frac{V+52}{7.4}}}, \quad \tau_{mT} = 0.44 +\frac{0.15}{e^{\frac{V+27}{10}}+e^{-\frac{V+102}{15}}}\\

% \dfrac{dh_T}{dt} = \frac{h_T^\infty-h_{T}}{\tau_{hT}}, \quad h_{T}^\infty = \frac{1}{1+e^{\frac{V+80}{5}}}, \quad \tau_{hT} = 62.7+\frac{0.27}{e^{\frac{V+48}{4}}+e^{-\frac{V+407}{50}}}\\

% \dfrac{dCa}{dt} = -\frac{kI_T}{2Fd}-\frac{K_TCa}{Ca+K_d},
% \end{array}
% \end{equation}
\begin{equation}
\begin{cases}
\vspace{5pt}
\dfrac{dV}{dt} = \dfrac{-I_T-I_L-I_{Na}-I_{K}-I_c+I^{syn}}{C} \\\vspace{5pt}
\dfrac{dCa}{dt} = -\dfrac{kI_T}{2Fd}-\dfrac{K_TCa}{Ca+K_d} \\
\dfrac{dy}{dt} = \dfrac{y^\infty -y}{\tau_y}     \quad y = \{h,m,n,m_T,h_T\}
\end{cases}
\end{equation}
\noindent where $V$ is the membrane potential of the neuron; the ion currents $I_T$ (calcium), $I_{Na}$ (sodium), $I_K$ (potassium), and $I_L$ (leakage)
evolve according to the following equations
\begin{align*}
&I_T = g_{Ca}m_T^2h_T(V-E_{Ca}), \quad I_L = g_L(V-E_L), \\
&I_{Na} = g_{Na}m^3h(V-E_{Na}), \quad I_K = g_kn^4(V-E_k),
\end{align*}
which depend on $V$, on the intracellular calcium concentration $Ca$ and on a set of further state variables (called \textit{gating variables}) $h$, $m$, $n$, $m_T$, $h_T$. The differential equations governing these gating variables have the common structure written above (for the generic gating variable $y$), where:
\begin{align*}
&y^\infty = a_y/(a_y + b_y), \quad \tau_y = 1/(a_y + b_y) \quad (y = \{h,m,n\})\\
&a_h = 0.128e^{\frac{17-V}{18}}, \quad b_h = \frac{4}{e^{-0.2(V-40)}+1}, \\
&a_m = \dfrac{0.32(13-V)}{e^{0.25(13-V)}-1}, \quad b_m = \frac{0.28(V-40)}{e^{0.2(V-40)}-1}\\
&a_n = \dfrac{0.032(15-V)}{e^{0.2(15-V)}-1}, \quad b_n = 0.5e^{\frac{10-V}{40}}\\
&m_{T}^\infty = \frac{1}{1+e^{-\frac{V+52}{7.4}}}, \quad \tau_{mT} = 0.44 +\frac{0.15}{e^{\frac{V+27}{10}}+e^{-\frac{V+102}{15}}}, \\
&h_{T}^\infty = \frac{1}{1+e^{\frac{V+80}{5}}}, \quad \tau_{hT} = 62.7+\frac{0.27}{e^{\frac{V+48}{4}}+e^{-\frac{V+407}{50}}} .
\end{align*}

In the above equations, $h$ and $m$ are the inactivation and activation variables of the $Na^+$ current; $n$ is the activation variable of the $K^+$ current; $m_T$ and $h_T$ are the activation and inactivation variables of the low-threshold $Ca^{2+}$ current; the leakage current $I_L$ has conductance $g_L = 0.05 \,[\frac{mS}{cm^2}]$ and reversal potential $E_L = -78 \,[mV]$; $I_{Na}$ and $I_K$ are the fast $Na^+$ and $K^+$ currents responsible for the generation of action potentials, with conductances $g_{Na} = 100 \,[\frac{mS}{cm^2}]$ and $g_k = 10 \,[\frac{mS}{cm^2}]$ and reversal potentials $E_{Na} = 50 \,[mV]$ and $E_k = -95 \,[mV]$; $I_T$ is the low-threshold $Ca^{2+}$ current that mediates the rebound burst response, with conductance $g_{Ca} = 1.75 \,[\frac{mS}{cm^2}]$ and reversal potential $E_{Ca}=k_0\frac{RT}{2F}\log(\frac{Ca_0}{Ca})$; $I^{syn}$ is the synaptic current (Eq. (2) in the paper).

When the control current $I_c$ is in the range $\,[-0.43,0.13]\,\,[\frac{\mu A}{cm^2}]$ the neuron exhibits bursting behavior.
The other parameters are set as follows:
$C = 1 \,[\frac{\mu F}{cm^2}], \quad Ca_0 = 2 \,[mM], \quad d = 1 \,[\mu m]$, $K_T = 0.0001 \,[mM\cdot ms],\quad K_d = 0.0001 \,[mM]$. $F=96.489 \,[\frac{C}{mol}]$ is the Faraday constant, $R=8.31441\,[\frac{J}{mol\cdot K}]$ is the universal gas constant and the temperature $T$ is set at $309.15 \,[K]$.

\subsection{Exponential Integrate and Fire Neuron Model}
The exponential integrate and fire (eIF) neuron model \cite{brette2005adaptive} is defined by the following state equations:
\begin{equation}
\begin{cases}
    \dfrac{dV}{dt}=\dfrac{-g_L(V-E_L)+g_ee^{\frac{V-V_T}{\Delta_T}}-u+I_{ext}+I^{syn}}{C} \\
    \\
    \dfrac{du}{dt}=\dfrac{a(V-E_L)-u}{\tau_w}
\end{cases}
\end{equation}
where $V$ is the membrane potential of the neuron; $u$ is the adaptation variable; $g_L = 30 \,[nS]$ is the leakage conductance and $E_L = -70.6 \,[mV]$ is the leakage reversal potential; $I^{syn}$ is the synaptic current (Eq. (2) in the paper).

When the conductance $g_e$ is set at $110\,[nS]$, the external current $I_{ext}$ is varied in the range $[690,1110]\,[pA]$ (Fig. \ref{fig:condition_C}). When the external current $I_{ext}$ is set at $800\,[pA]$, the conductance $g_e$ is varied in the range $[20,160]\,[nS]$ (Fig. \ref{fig:support}). For this range of parameter values, the neuron exhibits bursting behavior.
The other parameters are set as follows: $C = 2007.4 \,[pF],\quad V_T = -50.4 \,[mV],\quad \Delta_T = 2\,[mV],\quad \tau_w = 285.7\,[ms],\quad a = 4\,[nS]$.

\subsection{Plant Neuron Model}
 The Plant neuron model \cite{plant1981bifurcation,alaccam2015making} is defined by the following state equations:
\begin{equation}
\begin{cases}
\vspace{5pt}
\dfrac{dV}{dt} = \dfrac{-I_{T}-I_L-I_{Na}-I_K-I_{KCa}+I_{ext}+I^{syn}}{C} \\
\dfrac{dCa}{dt} = \rho (K_cx(V_{Ca}-V)-Ca)\\
\\
\dfrac{dy}{dt} =  \dfrac{y^{\infty}-y}{\tau_y}  \quad y = \{h,n,x\}
\end{cases}
\end{equation}
where
\begin{align*}
&I_T = g_Tx(V-E_I), \quad I_L = g_L(V-E_L), \\
&I_{Na} = g_Im_{\infty}^3h(V-E_I), I_K = g_Kn^4(V-E_K),\\
&I_{KCa} = g_{KCa}\frac{Ca}{Ca+0.5}(V-E_K), \\
&m_{\infty}=\frac{\frac{0.1(50-V_s)}{e^{\frac{50-V_s}{10}}}}{\frac{0.1(50-V_s)}{e^{\frac{50-V_s}{10}}}+4e^\frac{25-V_s}{18}},\\ 
&h^{\infty}=\frac{0.07e^{\frac{25-V_s}{20}}}{0.07e^{\frac{25-V_s}{20}}+\frac{1}{1+e^\frac{55-V_s}{10}}}, \\ 
&\tau_h=\frac{12.5}{0.07e^{\frac{25-V_s}{20}}+\frac{1}{1+e^\frac{55-V_s}{10}}}, \\
&n^{\infty}= \frac{\frac{0.01(55-V_s)}{e^{\frac{55-V_s}{10}}}-1}{\frac{0.01(55-V_s)}{e^{\frac{55-V_s}{10}}}-1+0.125e^{\frac{45-V_s}{80}}}, \\ &\tau_n=\frac{12.5}{\frac{0.01(55-V_s)}{e^{\frac{55-V_s}{10}}}-1+0.125e^{\frac{45-V_s}{80}}}, \\
&x^{\infty}=\frac{1}{e^{0.15(-V-50)}+1}, \\
&V_s=\frac{127V}{105}+\frac{8265}{105}
\end{align*}
where $V$ is the membrane potential of the neuron; $Ca$ is the intracellular calcium concentration; $x$ is the activation variable of the slow inward $Ca^{2+}$ current; $h$ is the inactivation variable of the $Na^+$ current; $n$ is the activation variable of the $K^+$ current;
% and $y$ is the inactivation variable of the fast inward current.
% &m_{\infty}=\frac{0.1(50-\frac{124V+8265}{105})}{(e^{\frac{50-\frac{127V+8265}{105}}{10}}-1)(\frac{0.1(50-\frac{127V+8265}{105})}{e^{\frac{50-\frac{127V+8265}{105}}{10}}-1}+4e^{\frac{25-\frac{127V+8265}{105}}{18}})}
$I_L$ is the leakage current, with conductance $g_L = 0.003 \,[nS]$ and reversal potential $E_L = -40\,[mV]$; $I_{Na}$ and $I_K$ are the fast inward $Na^+$ and outward $K^+$ currents, respectively, with conductances $g_I = 8\,[nS]$ and $g_K = 1.3\,[nS]$ (these values ensure undershoot, see paper) and reversal potentials $E_I = 30mV$ and $E_K = -75\,[mV]$; $I_T$ is the slow inward tetrodotoxin-resistant $Ca^{2+}$ current, with conductance $g_T = 0.01\,[nS]$ and reversal potential $E_T=30\,[mV]$; $I_{KCa}$ is the outward $Ca^{2+}$ sensitive $K^+$ current, with conductance $g_{KCa}=0.03\,[nS]$ and reversal potential $E_K$; $I^{syn}$ is the synaptic current (Eq. (2) in the paper).
% $I_h$ is the fast inward current, defined by the conductance $g_h=0.001nS$ and the reversal potential $E_h=120mV$.

The external current $I_{ext}$ is set to $0.028 \,[\mu A]$.
The other parameters are set as follows: $C = 1 \,[\frac{\mu F}{cm^2}],\quad \rho=0.00015 \,[mV^{-1}],\quad K_c=0.0085\,[mV^{-1}],\quad V_{Ca}=140\,[mV],\quad \tau_x=235\,[ms]$.

\section{Synapse Parameter Values}
In Table \ref{tab:parameters}, column A lists the parameter values used for the gHCO with the thalamic reticular neuron model, first-order dynamic excitatory synapses and static inhibitory synapses (Figs. \ref{fig:exc_inh_act} and \ref{fig:bif_diag_s-alpha}). Column B lists the parameter values used when simulating the gHCO with the thalamic reticular neuron model, modified first-order dynamic inhibitory synapses (Eq. \ref{eq:alpha2}) and static excitatory synapses (Figs. \ref{fig:bifurcation_swap} and \ref{fig:smooth} in the paper). Column C lists the parameter values used for the gHCO with the eIF neuron model when varying $I_{ext}$, first-order dynamic excitatory synapses and static inhibitory synapses (Fig. \ref{fig:condition_C}). Column E lists the parameter values used for the gHCO with the eIF neuron model when varying $g_e$, first-order dynamic excitatory synapses and static inhibitory synapses (Fig. \ref{fig:support}). Column E lists the parameter values used for the gHCO with the Plant neuron model, first-order dynamic excitatory synapses and static inhibitory synapses (Fig. \ref{fig:condition_B}). 

% \begin{table}[h!]
% \caption{\label{tab:parameters}
% Parameter values.}
% \begin{ruledtabular}
% \begin{tabular}{ccccc}
% &A&B&C&D\\
% \hline
% $a^e [ms^{-1}]$& 0.5 & - & 50 & 0.2\\
% $b^e [ms^{-1}]$& 0.02 & - & 0.1 & 0.001\\
% $\theta^e [mV]$& 10 & -30 & -40 & -20\\
% $g^e [\frac{mS}{cm^2}]$& 0.0005 & 0.00001 & 1 & 0.00001\\
% $E^e [mV]$& 60 & 60 & 20 & 20\\
% $a^i [ms^{-1}]$& - & 0.5 & - & -\\
% $b^i [ms^{-1}]$& - & 0.02 & - & -\\
% $\theta^i [mV]$& -30 & 25 & -48.5 & -54\\
% $g^i [\frac{mS}{cm^2}]$& 0.0005 & 0.01 & 1 & 0.00001\\
% $E^i [mV]$& -80 & -80 & -100 & -55\\
% $\nu [mV^{-1}]$& 10 & 10 & 10 & 10\\
% \end{tabular}
% \end{ruledtabular}
% \end{table}

% \begin{table}[h!]
% \caption{\label{tab:parameters}
% Parameter values.}
% \begin{ruledtabular}
% \begin{tabular}{ccccc}
% &A&B&C&D\\
% \hline
% $a^e $& 0.5 & - & 50 & 0.2\\
% $b^e $& 0.02 & - & 0.1 & 0.001\\
% $\theta^e $& 10 & -30 & -40 & -20\\
% $g^e $& 0.0005 & 0.00001 & 1 & 0.00001\\
% $E^e $& 60 & 60 & 20 & 20\\
% $a^i $& - & 0.5 & - & -\\
% $b^i $& - & 0.02 & - & -\\
% $\theta^i $& -30 & 25 & -48.5 & -54\\
% $g^i $& 0.0005 & 0.01 & 1 & 0.00001\\
% $E^i $& -80 & -80 & -100 & -55\\
% $\nu $& 10 & 10 & 10 & 10\\
% \end{tabular}
% \end{ruledtabular}
% \end{table}

\begin{table}[h!]
\caption{\label{tab:parameters}
Parameter values.}
\begin{ruledtabular}
\begin{tabular}{cccccc}
&A&B&C&D&E\\
\hline
$\alpha^{ex} $& 0.1556 & - & 10 & 10 & 0.5\\
$\beta^{ex} $& 0.005 & - & 0.26 & 26 & 0.0005\\
$\theta^{ex} $& 25 & -30 & -40 & -40 & -42\\
$g^{ex} $& 0.0005 & 0.00001 & 1 & 0.4 & 0.0001\\
$E^{ex} $& 60 & 60 & 20 & 20 & 50\\
$\alpha^{in} $& - & 0.5 & - & - & -\\
$\beta^{in} $& - & 0.02 & - & - & -\\
$\theta^{in} $& -30 & 25 & -48.5 & -48.5 & -53\\
$g^{in} $& 0.0005 & 0.01 & 0.6 & 0.1 & 0.0001\\
$E^{in} $& -80 & -80 & -110 & -110 & -80\\
$\nu $& 10 & 10 & 10 & 10 & 10\\
\end{tabular}
\end{ruledtabular}
\end{table}

%\vspace{3cm}}\\

\providecommand{\noopsort}[1]{}\providecommand{\singleletter}[1]{#1}%

\end{document}